\documentclass[12pt]{article}
\usepackage{amsfonts}
\usepackage{graphicx}
\usepackage{epstopdf}
\usepackage{epsfig}
\usepackage{amssymb}
\usepackage{setspace}
\usepackage{caption}
\usepackage{amsfonts}
\usepackage{color}
\usepackage{amsmath}
\usepackage{float}
\usepackage{comment}
\newcommand {\be}{\begin{equation}}
\newcommand {\ee}{\end{equation}}
\newcommand {\bea}{\begin{array}}
	
	\newcommand {\eea}{\end{array}}

\newcommand {\RN}{Reissner–Nordstrom~}

\textwidth=6.5in \hoffset=-.75in \voffset=-1in \numberwithin{equation}{section}
\numberwithin{figure}{section}

\begin{document}

	\begin{titlepage}
		\vspace{1cm}
		\begin{center}
			{\Large \bf {Weakly magnetized black holes in Einstein-ModMax theory}}\\
		\end{center}
		\vspace{2cm}
		\begin{center}
			\renewcommand{\thefootnote}{\fnsymbol{footnote}}
			Haryanto M. Siahaan{\footnote{haryanto.siahaan@unpar.ac.id}}\\
Universitas Katolik Parahyangan,\\
			Jalan Ciumbuleuit 94, Bandung 40141, Indonesia
			\renewcommand{\thefootnote}{\arabic{footnote}}
		\end{center}
		
		\begin{abstract}
			
Theories of non-linear electrodynamics inherently describe deviations from Maxwell’s theory in the strong field regime. Among these, ModMax electrodynamics stands out as a unique one-parameter generalization of Maxwell’s theory that preserves both conformal invariance and electromagnetic duality. In this paper, we investigate the extension of Wald's magnetization within the framework of Einstein-ModMax theory, concentrating on static charged and accelerating black holes. Additionally, we examine the influence of external magnetic fields on the motion of charged test particles in the vicinity of a charged black hole.
			
		\end{abstract}
	\end{titlepage}\onecolumn
	\bigskip

\section{Introduction}\label{sec.intro}
\label{sec:intro}

Black holes are an inevitable implication of Einstein's theory of relativity, predicting their existence long before they were finally observed about a century later \cite{EventHorizonTelescope:2019dse}. In Einstein-Maxwell theory, black holes can possess electric charge in addition to mass and rotational parameters. One of the most renowned exact solutions describing a rotating and electrically charged black hole in this theory is the Kerr-Newman solution \cite{Novikov:1989sz}. Beyond mass, rotation, and charge, exact solutions in Einstein-Maxwell theory can also include NUT and acceleration parameters. The Plebanski-Demianski solution \cite{Plebanski:1976gy} is one of the most comprehensive black hole solutions in Einstein-Maxwell theory, encompassing multiple parameters.

Astrophysical black holes are often surrounded by external magnetic fields. Recent astronomical observations have revealed strong, organized magnetic fields spiraling from the edge of the supermassive black hole Sagittarius A* \cite{Collaboration:2024unf}. Previous studies by the Event Horizon Telescope Collaboration have also detected strong external magnetic fields around black holes at the centers of galaxies, inferred through the observed light polarization \cite{EventHorizonTelescope:2021bee,EventHorizonTelescope:2021srq}. The direction of circular light polarization, whether clockwise or counterclockwise, as it travels, provides information about the magnetic field and the types of high-energy particles surrounding the black hole. These findings have renewed interest in applying models of magnetized black holes to study the motion of objects around them.

In Einstein-Maxwell theory, there are two common approaches to modeling the interaction between a black hole and an external magnetic field. The first approach treats the external magnetic field as a perturbation in spacetime, constructing the associated vector potentials based on the existing Killing vectors of the spacetime \cite{Wald:1974np}. The second approach involves using the Ernst transformation on a seed solution \cite{Ernst:1976mzr}, which can be any member of the Kerr-Newman-Taub-NUT family, to incorporate the influence of the magnetic field on spacetime curvature \cite{Siahaan:2021uqo,Ghezelbash:2021lcf}. This latter approach can be used to investigate how the external magnetic field modifies the null geodesics around a black hole.

In recent years, there has been growing interest in a particular type of non-linear electrodynamics known as modified Maxwell (ModMax) theory \cite{Bandos:2020hgy}. The appeal of studying non-linear electrodynamics lies in its potential to address the issue of field singularities. Various aspects of ModMax electrodynamics and the corresponding black hole solutions have been extensively investigated, as seen in works such as \cite{Sorokin:2021tge, Kosyakov:2020wxv, Bandos:2020hgy, Bandos:2021rqy, Kruglov:2021bhs, Flores-Alfonso:2020euz, BallonBordo:2020jtw, Kubiznak:2022vft, Barrientos:2022bzm, Siahaan:2023gpc, Siahaan:2024ajn, EslamPanah:2024fls, EslamPanah:2024tex}. In ModMax theory, the non-linear parameter effectively "screens" the charges through an exponential factor. Interestingly, the exact solution for a static charged black hole in Einstein-ModMax theory closely resembles the well-known Reissner-Nordstrom solution \cite{Flores-Alfonso:2020euz}. Aspects of a dyonic Einstein-ModMax black hole, such as shadow, lensing, and quasinormal modes, have been investigated in \cite{Pantig:2022gih}, with the potential for these phenomena to be detected as astronomical observational equipment continues to improve.

With the concepts of magnetized black holes and Einstein-ModMax theory in mind, a natural question arises: what if we consider magnetization within the framework of Einstein-ModMax theory, at least at the perturbative level as introduced by Wald \cite{Wald:1974np}? This approach is simpler compared to the more complex task of constructing an exact magnetized black hole solution in Einstein-ModMax theory, as done by Ernst in the Einstein-Maxwell setup \cite{Ernst:1976mzr}. If a magnetized black hole in Einstein-ModMax theory can be constructed following Wald's method, we can study the motions of charged timelike objects to understand how the external magnetic field and non-linear parameter induce deviations. This paper aims to do precisely that: construct a magnetized black hole in Einstein-ModMax theory and examine the resulting motions of charged timelike objects. 
To the best of our knowledge, this is the first study to apply Wald's magnetization in the context of Einstein gravity with non-linear electrodynamics. Note that, there exist a recent work reported in \cite{Barrientos:2024umq} where the authors constructed electromagnetized black holes and vortex-like backgrounds within the framework of the ModMax theory.

The organization of this paper is as follows. In the next section, we provide a brief review of Einstein-ModMax theory and the corresponding static charged black hole solution. Section \ref{sec.WeaklyMag} discusses the construction of a magnetized black hole in Einstein-ModMax theory. In section \ref{sec.Motions}, we examine the motions of charged timelike objects in the magnetized charged black hole. Finally, we present our conclusions. In this paper, we consider the natural units where $c={\hbar} = k_B = G_4 = 1$.

\section{Charged black holes in Einstein-ModMax theory}\label{sec.EinsteinModMax}

The ModMax theory \cite{Bandos:2020hgy,Kosyakov:2020wxv} is described by Lagrangian density
\be 
{\cal L}_{MM}  =  - \frac{1}{2}\left( {s\cosh v - \sqrt {s^2  + p^2 } \sinh v} \right)
\ee 
where $s$ and $p$ are the invariants of the electromagnetic fields, namely
\be \label{eq.sp}
s = \frac{1}{2}F_{\mu \nu } F^{\mu \nu} ~~,~~ p = \frac{1}{2}F_{\mu \nu } \tilde F^{\mu \nu } \,.
\ee 
In the equations above, the field strength tensor is defined as $F_{\mu \nu }  = \partial _\mu  A_\nu   - \partial _\nu  A_\mu  $ and its dual as $\tilde F_{\mu \nu }  = \frac{1}{2}\varepsilon _{\mu \nu \alpha \beta } F^{\alpha \beta } $ where ${\varepsilon _{0123} } = \sqrt { - g} $. In differential form notation, we have ${\bf {\tilde F}} = \star {\bf  F}$, where $\star$ denotes the Hodge dual star operator. The parameter $v$ denotes the non-linear parameter of the theory, with the standard Maxwell theory recovered when $v=0$. It has been shown that the condition $v \ge 0$ must be imposed to ensure causality \cite{Sorokin:2021tge}. Furthermore, since the ordinary Maxwell theory describes our nature extremely well, we can expect the parameter $v$ to be extremely small. 

Following \cite{BallonBordo:2020jtw, Barrientos:2022bzm}, we introduce the two-form for the "material" field strength as follows
\be \label{eq.TensorMaterial}
{\bf E} = 2\left( {f_s {\bf F} + f_p {\bf {\tilde F}}} \right)
\ee 
where ${\bf F} $ and ${\bf {\tilde F}}$ are the two-forms for the Maxwell field-strength tensor and its dual, respectively. The functions $f_s$ and $f_p$ are defined as
\be 
f_s  = \frac{{\partial {\cal L}_{MM} }}{{\partial s}}~~,~~f_p  = \frac{{\partial {\cal L}_{MM} }}{{\partial p}} \,.
\ee 
In terms of the functions above, the electromagnetic stress-energy tensor in ModMax theory can be written as
\be \label{eq.Tmn}
T_{\mu \nu }  = \frac{1}{{4\pi }}\left( {s~ g_{\mu \nu }  - 2 F_{\mu \kappa } F_{\nu \lambda } g^{\kappa \lambda } } \right) f_s \,.
\ee 
Furthermore, the electric charge inside a closed two dimensional spacelike surface $\Sigma $ can be obtained from the integral
\be \label{eq.Qe}
Q_e  = \frac{1}{{4\pi }}\int_\Sigma  {  \star {\bf  E}} \,.
\ee 

Now let us consider an action for Einstein-ModMax theory \cite{Flores-Alfonso:2020euz,BallonBordo:2020jtw,Barrientos:2022bzm}
\be 
S = \frac{1}{{16\pi }}\int {d^4 x\sqrt { - g} } \left( {R - 4{\cal L}_{MM} } \right) \,.
\ee 
From the action above, the corresponding equations of motion in Einstein-ModMax theory are
\be \label{eq.EinsteinEq}
R_{\mu \nu }  - \frac{1}{2}g_{\mu \nu }  = 8\pi T_{\mu \nu } \,,
\ee 
and
\be \label{eq.source-free}
\nabla ^\mu  E_{\mu \nu }  = 0\,.
\ee 
Note that the last equation represents a generalization of the source-free condition $\nabla ^\mu  F_{\mu \nu }  = 0$ found in ordinary Einstein-Maxwell theory. Despite the complexity of the corresponding equations of motion, it turns out that one of the simplest static black hole solutions describing a collapsed charged mass closely resembles the well-known Reissner-Nordstrom solution. The metric reads \cite{Flores-Alfonso:2020euz} 
\be \label{eq.Metric}
ds^2  =  - \left( {1 - \frac{{2M}}{r} + \frac{{e^{ - v} Q^2 }}{{r^2 }}} \right)dt^2  + \frac{{dr^2 }}{{\left( {1 - \frac{{2M}}{r} + \frac{{e^{ - v} Q^2 }}{{r^2 }}} \right)}} + r^2 \left( d\theta^2 + \sin^2\theta d\phi ^2  \right)\,,
\ee 
whereas the corresponding vector potential is given by
\be \label{eq.Anonmag}
A_\mu  dx^\mu   = -\frac{{e^{ - v} Q}}{r}dt\,.
\ee 
In equations above, $M$ and $Q$ represent the black hole mass and electric charge, respectively, whereas $v$ denotes the non-linear parameter of the ModMax theory.

For the solution described by equations (\ref{eq.Metric}) and (\ref{eq.Anonmag}), the non-vanishing components of the field strength tensor and its dual are given by ${\bf F} = -\frac{{e^{ - v} Q}}{{r^2 }}dt \wedge dr$
and ${\bf \tilde F} =   e^{ - v} Q\sin \theta dt \wedge dr$, respectively. Accordingly, by using eq. (\ref{eq.sp}), we can find 
\be 
s =  - \frac{{e^{ - 2v} Q^2 }}{{2r^4 }}\,,
\ee 
and $p=0$. Furthermore, the material tensor (\ref{eq.TensorMaterial}) for this solution can be read as
\be 
{\bf E} =   \frac{Q}{{r^2 }}dt \wedge dr \,,
\ee 
and $\star {\bf E} = Q\sin \theta d\theta  \wedge d\phi$. Therefore, the electric charge of the black hole can be computed using equation (\ref{eq.Qe}), yielding $Q_e = Q$.  

\section{Weakly magnetized black holes in ModMax theory}\label{sec.WeaklyMag}

In \cite{Wald:1974np}, magnetized black holes are studied by introducing external perturbative magnetic fields that are proportional to the Killing vectors of the spacetime. The Killing vectors $\zeta ^\mu  $ satisfy the Killing equations $\nabla ^\nu  \zeta ^\mu   + \nabla ^\mu  \zeta ^\nu   = 0$. Consequently, one can construct a field-strength tensor based on these Killing vectors, given by $
F_{\mu \nu }  = \nabla _\mu  \zeta _\nu   - \nabla _\nu  \zeta _\mu  $. This construction ensures that the source-free condition 
\be \label{eq.sourefreeEM}
\nabla _\mu  F^{\mu \nu }  = 0 
\ee 
is satisfied. For the Reissner-Nordstrom-ModMax solution reviewed in the previous section, there are two Killing vectors associated with the stationary and axial symmetries of the spacetime, i.e. $\zeta_{\left(t\right)}^\mu = \left[1,0,0,0\right]$ and  $\zeta_{\left(\phi\right)}^\mu = \left[0,0,0,1\right]$, respectively. For each Killing vector, we can associate a field strength tensor, namely ${\bf F}_{\left(t\right)}  \propto  d{\boldsymbol \zeta} _{\left( t \right)} $ and ${\bf F}_{\left(\phi\right)}  \propto  d{\boldsymbol \zeta} _{\left( \phi \right)} $, along with their corresponding dual tensors.

Let us now apply Wald's prescription to magnetize a black hole in Einstein-ModMax theory, and consider the generalized approach introduced by \cite{Azreg-Ainou:2016tkt}. The process begins similarly by introducing vector potentials based on the Killing vectors of the spacetime. However, in contrast to the Einstein-Maxwell theory, where vector potentials must satisfy the source-free condition given by equation (\ref{eq.sourefreeEM}), in Einstein-ModMax theory, the vector potentials must fulfill the generalized source-free condition as specified in equation (\ref{eq.source-free}). Consequently, the vector potential for a weakly magnetized black hole in Einstein-ModMax theory is governed by more stringent equations. For instance, while in ordinary Einstein-Maxwell theory, one can use the vector potential $A^\mu = \left[0,0,0,1\right]$ to construct the external magnetic field in describing the magnetized Kerr black hole \cite{Wald:1974np}, this approach is not applicable in the Einstein-ModMax theory for the Kerr solution due to the requirements of equation (\ref{eq.source-free}).

Continuing from the previous discussion, for the magnetized black hole in Einstein-ModMax theory, rather than directly using a constant test field proportional to the Killing vector, we adopt a more general approach based on the ansatz for vector potentials introduced by Azreg-A\"{\i}nou in \cite{Azreg-Ainou:2016tkt}. This approach allows us to explore a broader class of vector potentials, 
\be \label{eq.ACt}
A_{\left( t \right)}^\mu   = \left[ {C_t \left( r \right),0,0,0} \right] \,,
\ee 
and
\be \label{eq.ACp}
A_{\left( \phi \right)}^\mu   = \left[0,0,0, {C_\phi \left( r \right)} \right]\,.
\ee 
Thus, the test field corresponding to $\zeta_{\left(t\right)}^\mu$ can be computed as ${\bf F}_{\left(t\right)} =  d{\boldsymbol A} _{\left( t \right)}$, while the test field associated with $\zeta_{\left(\phi\right)}^\mu$ is given by ${\bf F}_{\left(\phi\right)} =  d{\boldsymbol A} _{\left( \phi \right)}$. Furthermore, the material tensor can be defined associated to each Killing vector as well, i.e. ${\bf E}_{\left(t\right)}$ and ${\bf E}_{\left(\phi \right)}$. For the test field constructed from ${\bf \zeta}_{\left(t\right)}$, the condition $d \star {\bf E}_{\left(t \right)} = 0$ results in an equation $C_t\left(r\right)$, namely
\be 
r^2 e^v \left( {r^2  - 2Mr + e^{ - v} Q^2 } \right)\frac{{d^2 C_t }}{{dr^2 }} - 2re^v \left( {r^2  - e^{ - v} Q^2 } \right)\frac{{dC_t }}{{dr}} - 2Q^2 C_t  = 0\,.
\ee 
A general solution to this equation can be derived, resembling the solution for a weakly magnetized Reissner-Nordstrom black hole discussed in \cite{Azreg-Ainou:2016tkt}. Specifically, it takes the form
\be 
C_t  = \frac{{C_1 }}{{rg_{tt} }} + \frac{{C_2 }}{{g_{tt} }}
\ee 
for the metric function as appeared in eq. (\ref{eq.Metric}). Finally, motivated by the vector solution in the non-magnetized case (\ref{eq.Anonmag}), we can take $C_2=0$ and $C_1 = -e^{ - v} Q$. 

On the other hand, for the test field associated to the axial Killing vector, equation $d \star {\bf E}_{\left(\phi\right)}$ gives
\be 
- r^2 e^v \left( {r^2  - 2Mr + e^{ - v} Q^2 } \right)\frac{{d^2 C_t }}{{dr^2 }} - 2re^v \left( {2r^2  + e^{ - v} Q^2  - 3Mr} \right)\frac{{dC_\phi  }}{{dr}} + 2Q^2 C_\phi   = 0\,,
\ee 
whose general solution can be expressed as
\be 
C_\phi   = \frac{{C_3 }}{{r^2 }}\left( {e^v r^2  - Q^2 } \right) + \frac{{C_4 }}{{r^2 }}\left( {\left( {Q^2  - e^v r^2 } \right)\tanh ^{ - 1} \left( {\frac{{\left( {M - r} \right)Z}}{{e^v M^2  - Q^2 }}} \right) - \left( {M + r} \right)Z} \right)
\ee
with $Z=\sqrt{e^{2v}M^2 - e^v Q^2}$. Indeed, taking the limit $v\to 0$ reduces the last equation to the axial component of the vector potential for a magnetized Reissner-Nordström black hole, as discussed in \cite{Azreg-Ainou:2016tkt}. To prevent singularities in the ${\bf F}_{\left(\phi\right)}$ component at the horizon, we set $C_4=0$. Additionally, to ensure a constant magnetic field as $r\to \infty$, similar to the magnetized Reissner-Nordstrom black hole, we set $C_3 = e^{-v} B/2$. In summary, the vector components for the test field associated with the Killing vectors of a magnetized, static, electrically charged black hole in ModMax theory are given by 
\be \label{eq.At}
A_{\left( t \right)}^\mu   = \left[ -{\frac{{e^{ - v} Q}}{r},0,0,0} \right]\,,
\ee 
and
\be \label{eq.Ap}
A_{\left( \phi  \right)}^\mu   = \left[ {0,0,0,\frac{B}{2}\left( {1 - \frac{{e^{ - v} Q^2 }}{{r^2 }}} \right)} \right]\,,
\ee 
which reduce to the vector field in the magnetized \RN after setting $v=0$. Note that for an electrically neutral black hole, the external magnetic field vector potential is identical to that in ordinary Einstein-Maxwell theory \cite{Wald:1974np}. In other words, for the magnetized Schwarzschild black hole solution in Einstein-ModMax theory, it mirrors the solution found in the standard Einstein-Maxwell theory. This implies that the non-linearity of the theory becomes relevant only when an electric charge is present.

Let us examine the applicability of Wald's magnetization in the context of accelerating black holes within Einstein-ModMax theory. As detailed in \cite{Barrientos:2022bzm}, solutions for accelerating black holes in Einstein-ModMax theory have been explored. Note that Wald's approach to magnetization for accelerating black holes appears to be absent from the existing literature. Here let us just consider to magnetize a neutral accelerating black hole where the line element can be expressed as \cite{Hong:2003gx,Griffiths:2009dfa}
\be 
ds^2  = \frac{1}{\Omega^2}\left[ -G dt^2 + \frac{{dr^2 }}{{G} } + r^2 \frac{d\theta^2}{H} +r^2 H \sin^2 \theta d\phi^2  \right] \,,
\ee 
where $\Omega = 1+\alpha r \cos\theta$, $G= \left(r^2-2Mr\right) \left(1-\alpha^2 r^2\right)$, and $H=1+2M\alpha \cos\theta$. In these equations, $M$ and $\alpha$ represent the mass and acceleration parameters, respectively. The external magnetic field potentials can be constructed as constants multiplied by the corresponding Killing vectors ${\boldsymbol \zeta}{(t)}$ and ${\boldsymbol \zeta}{(\phi)}$, with the generalized source-free condition (\ref{eq.source-free}) remaining satisfied. This setup represents an accelerating black hole immersed in an external magnetic field according to the Einstein-ModMax theory. However, if one introduces an electric charge to the accelerating black hole as presented in \cite{Barrientos:2022bzm}, the generalized source-free condition (\ref{eq.source-free}) no longer holds. This situation mirrors the case in ordinary Einstein-Maxwell theory, where the vector potentials $A^\mu_{(t)} = [1,0,0,0]$ or $A^\mu_{(\phi)} = [0,0,0,1]$ also fail to satisfy the source-free condition (\ref{eq.sourefreeEM}).

Furthermore, we may consider magnetizing the Taub-NUT or Kerr black hole in Einstein-ModMax theory. To achieve this, we can employ the general vector potential
\be \label{eq.vector}
A^\mu = C_1 \zeta_{\left(t\right)}^\mu + C_2 \zeta_{\left(\phi\right)}^\mu
\ee 
where $C_1$ and $C_2$ are constants. Nevertheless, despite originating from relatively simple Einstein vacuum solutions, the generalized source-free condition (\ref{eq.source-free}) remains unsatisfied even for the vector potential given by (\ref{eq.vector}). In contrast, for the magnetized Kerr solution in Einstein-Maxwell theory, the vector potential (\ref{eq.vector}) satisfies the source-free condition (\ref{eq.sourefreeEM}). Moreover, when considering Kerr or Taub-NUT spacetimes, employing the vector potentials (\ref{eq.ACt}) and (\ref{eq.ACp}) that could potentially satisfy (\ref{eq.source-free}) leads to a highly complex differential equation for the vector function.

\section{Motions of a test charged timelike object under the influence of external magnetic fields}\label{sec.Motions}

To explore the astronomical implications of our previous studies, we now examine a charged timelike object as a perturbation in the spacetime of a charged, static ModMax black hole immersed in an external magnetic field described by the line element (\ref{eq.Metric}). To satisfy the generalized source-free condition (\ref{eq.source-free}), one must use either (\ref{eq.At}) or (\ref{eq.Ap}), but not both simultaneously. For our purposes, we will focus on the external electromagnetic potential given by
\be \label{eq.Amu}
A_\mu  dx^\mu   = \frac{B}{2}\sin ^2 \theta \left( {r^2  - e^{ - v} Q^2 } \right) d\phi\,.
\ee 
The Lagrangian for a charged timelike object with unit mass is given by
\be
{\cal L} = \frac{1}{2}g_{\mu \nu } \dot x^\mu  \dot x^\nu   + qA_\mu  \dot x^\mu \,,
\ee
where $q$ represents the charge per unit mass of the test object. Here, the notation ``dot'' denotes differentiation with respect to the affine parameter $\sigma$. The equatorial plane is defined by $\theta=0$, and the presence of equatorial motions can be analyzed as follows. On the equator, we have
\be \label{ELeqx}
\frac{{\partial {\cal L}}}{{\partial \theta}} - \frac{d}{{d\sigma }}\frac{{\partial {\cal L}}}{{\partial \dot \theta}} = 0 \,.
\ee 
Moreover, it can be verified for the metric (\ref{eq.Metric}) and vector potential (\ref{eq.Amu}) that
\be
\left. {\partial _\theta g_{\alpha \beta } } \right|_{\theta = \pi/2}  = 0~~,~~
\left. {\partial _x A_\alpha  } \right|_{\theta = \pi/2}  = \pi/2 \,.
\ee
It yields the equation (\ref{ELeqx}) reduces to
\be 
\frac{d}{{d\sigma }}\frac{{\partial {\cal L}}}{{\partial \dot \theta}} = 0\,.
\ee
It is evident that the final equation is met, as we are examining an object with no momentum and acceleration in the $\theta$ direction, i.e. $\dot \theta=0$ and ${\ddot \theta}=0$ at the equator. Consequently, we can infer the possibility of equatorial motion in the magnetized spacetime under consideration, with further deliberation on the circular motion forthcoming.

In this magnetized spacetime, which has both stationary and axial symmetry, a test object has two conserved quantities, namely the energy and angular momentum. The energy $E$ is given by
\be\label{eq.E}
E = - \frac{{\partial {\cal L}}}{{\partial \dot t}} = - {g_{tt} \dot t }\,,
\ee
where $\dot t$ and $\dot \phi$ represent the derivatives of the object's coordinate time and azimuthal angle, respectively, and $A_t$ is the time component of the electromagnetic potential. The angular momentum $L$ is given by
\be\label{eq.L}
L = \frac{{\partial {\cal L}}}{{\partial \dot \phi }} = g_{\phi \phi } \dot \phi - qA_\phi\,.
\ee
Equatorial motions demand $\dot \theta=0$, thence we can obtain a general equation from the metric, given by $g_{tt} \dot t^2 + g_{\phi \phi } \dot \phi ^2 + 1 = - g_{rr} \dot r^2 $. To study the circular motions, we can introduce an effective potential $V_{\rm eff} = - {\dot r}^2$. By plugging in the conserved quantities in eqs. (\ref{eq.E}) and (\ref{eq.L}), we obtain the expression for the effective potential as
\be 
V_{{\rm{eff}}}  = \frac{1}{{r^2 }}\left\{ {\frac{{\Delta _r }}{{r^2 }}\left[ {L + \frac{{qB}}{2}\left( {r^2  - e^{ - v} Q^2 } \right)} \right]^2  + \Delta _r  - r^2 E^2 } \right\}
\ee 
where $\Delta_r = r^2  - 2Mr + e^{ - v} Q^2 $.
In equations above, $E$ and $L$ denote the energy and angular momentum, respectively, of the test object. It is important to note that, in the effective potential, the magnetic field parameter affects only charged particles. This contrasts with cases of strongly magnetized spacetimes, as discussed in \cite{Ernst:1976mzr,Siahaan:2021uqo,Ghezelbash:2021lcf}, where even the geodesics of null objects are influenced by the presence of an external magnetic field.

\begin{figure}[!htb]
	\centering 	\includegraphics[scale=0.35]{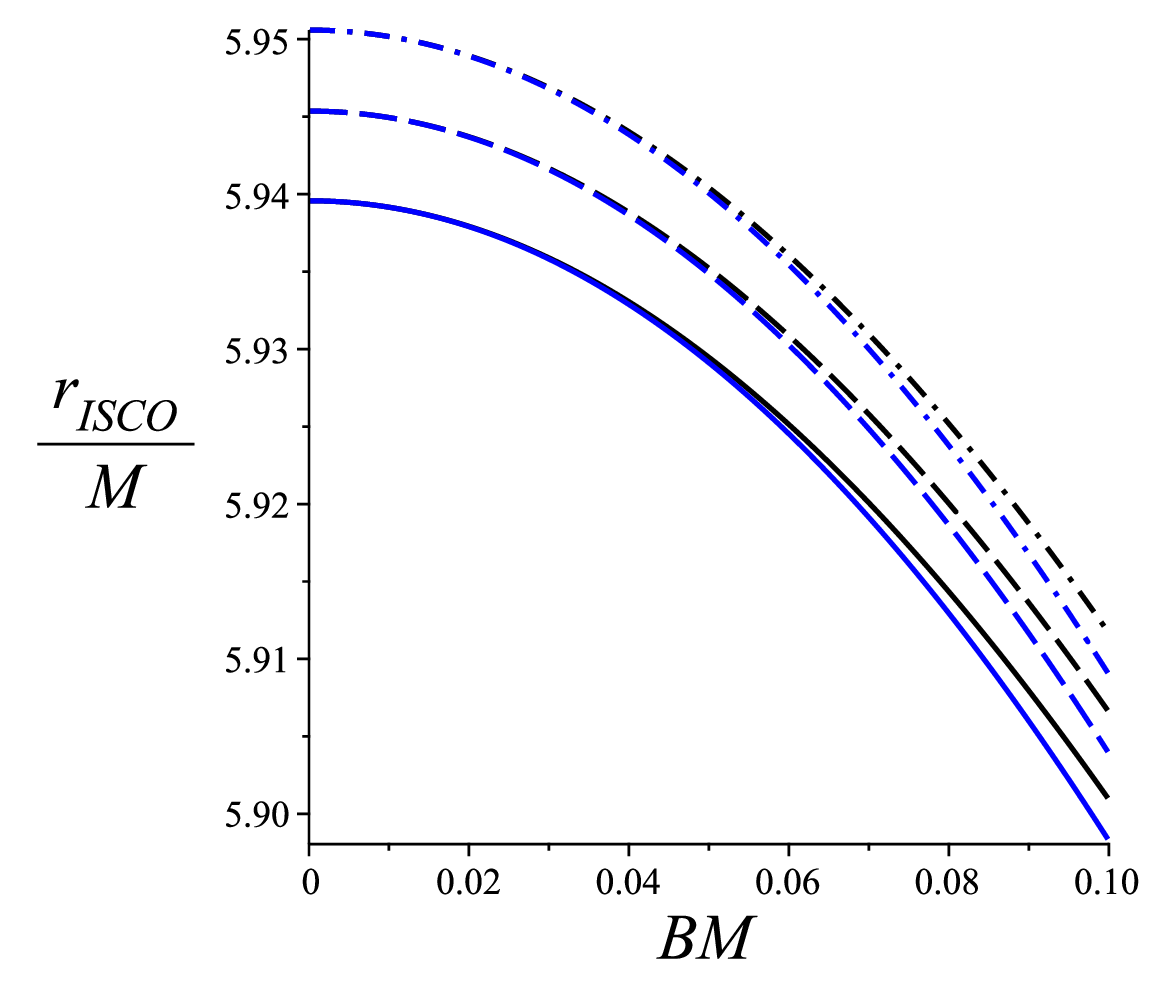}\caption{The ISCO radius of a charged timelike object in the presence of an external magnetic field is shown here. We consider parameters $Q=0.2 M$ and $q=0.1$. The black curves indicate Anti-Larmor orbits, while the blue curves represent Larmor orbits. The cases with $v=0$, $v=0.1$, and $v=0.2$ are depicted by solid, dashed, and dash-dotted curves, respectively.}\label{fig.rISCO}
\end{figure}

\begin{figure}[!htb]
	\centering 	\includegraphics[scale=0.35]{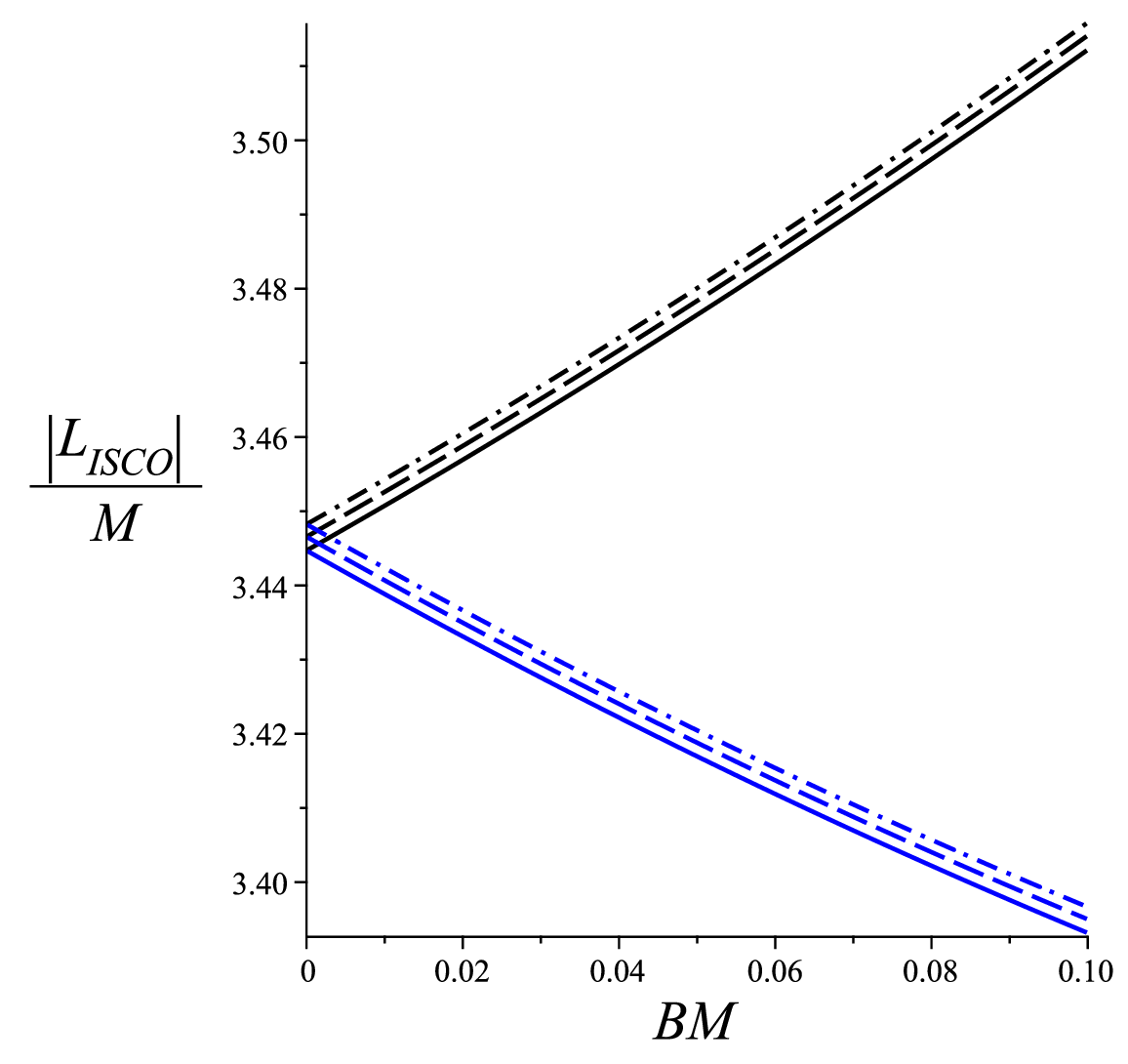}\caption{The absolute values of angular momentum for Larmor and Anti-Larmor motions for the test object illustrated in the plots in Fig. \ref{fig.rISCO}. Each object corresponds to a specific curve type and color.}\label{fig.LISCO}
\end{figure}

In the presence of a magnetic field in spacetime, the effective potential increases without bound as the radius becomes larger. This behavior aligns with the scenario of a magnetized spacetime, where the magnetic field extends even to asymptotic regions. Additionally, since the magnetic field couples to the particle's charge in the effective action, the singular values of the effective potential do not apply to neutral timelike objects. To determine the innermost stable circular orbits (ISCO) on the equatorial plane for a timelike object, we must solve the following system of equations simultaneously to find the ISCO radius
\be 
V_{{\rm{eff}}}  = 0~~,~~\frac{{dV_{{\rm{eff}}} }}{{dr}} = 0~~,~~\frac{{d^2 V_{{\rm{eff}}} }}{{dr^2 }} = 0\,.
\ee
By solving the three equations simultaneously, we determine the parameters $E$, $L$, and $r$ which represent the energy, angular momentum, and ISCO radius of a charged timelike object in circular motion. Notably, even though the magnetized black hole under consideration is static, the angular momentum $L$ can take both positive and negative real values. This phenomenon was also observed in \cite{Aliev:2002nw}, where positive angular momentum corresponds to an Anti-Larmor orbit and negative angular momentum represents a Larmor orbit. Larmor motion is associated with the Lorentz force directing the object towards the black hole, while Anti-Larmor motion is directed away. The numerical results presented in Figs. \ref{fig.rISCO} and \ref{fig.LISCO} illustrate these characteristics, showing that the ISCO radii for Larmor orbits tend to be smaller compared to those for Anti-Larmor orbits as the external magnetic field increases.

	\begin{figure}[!htb]
	\centering 	\includegraphics[scale=0.35]{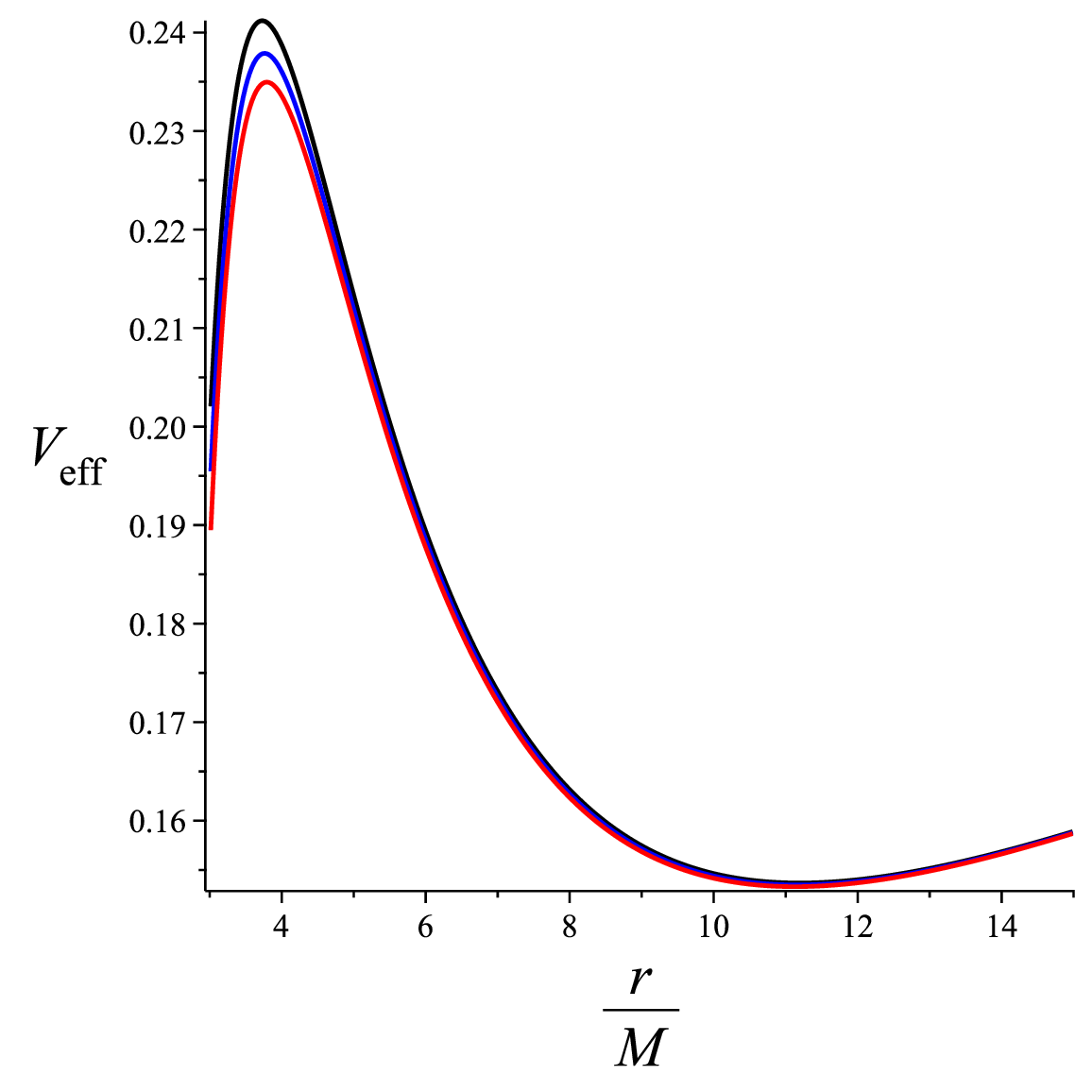}\caption{Effective potential for a charged test object with a charge-to-mass ratio $q=0.1$, energy $E=0.9$, black hole charge $Q=0.5M$, and magnetic field strength $BM=0.1$ in Einstein-ModMax theory, for various non-linear parameters $v$. The black, blue, and red curves correspond to $v=0.1$, $v=0.2$, and $v=0.3$, respectively.}\label{fig.Veffv}
\end{figure}

The impact of the non-linear parameter $v$ on circular motions is evident from the numerical examples for ISCO radii in Fig. \ref{fig.rISCO} and the effective potential in Fig. \ref{fig.Veffv}. It is observed that larger values of $v$ correspond to a lower peak of the local maxima in the effective potential. This indicates a screening effect due to the $v$ parameter,  meaning that the effective charge of the black hole decreases as $v$ increases.

\section{Conclusion}

In this work, we have demonstrated how to magnetize a black hole within the framework of Einstein-ModMax theory. We follow Wald's prescription by constructing the external magnetic field vector potential using the Killing vectors of the spacetime. Due to the complexity of the generalized source-free condition in Einstein-ModMax theory, not all magnetized black hole constructions from Einstein-Maxwell theory can be replicated in the ModMax context. For instance, the magnetized Kerr black hole in ModMax theory cannot be obtained using the ${\boldsymbol \zeta} _{\left( \phi \right)}$ Killing vector of the spacetime.

Nevertheless, we can achieve a magnetized static charged black hole in Einstein-ModMax theory even using the generalized prescription introduced in \cite{Azreg-Ainou:2016tkt}. As expected, the non-linear parameter in the vector potential solutions continues to play a role in screening the black hole's charge. To differentiate the motions of test charged objects for various non-linear parameter examples, section \ref{sec.Motions} presents numerical results related to the circular motions of the objects and the corresponding effective potentials. It was found that the general properties of the motions of charged objects in the magnetized ModMax black hole spacetime are similar to those in the magnetized black hole in Einstein-Maxwell theory \cite{Shaymatov:2021qvt}, with some discrepancies arising from the screening effect on the effective black hole charge.

For future work, we plan to study the generalized Wald magnetization of black holes in other non-linear Einstein-Maxwell theories, such as the Born-Infeld, Euler-Heisenberg, and Ayon-Beato solutions \cite{Cai:2004eh,Damour:1976jd,Ramadhan:2023ogm}. Additionally, during the preparation of this work, we discovered that equatorial circular motion for a timelike test object cannot exist in an accelerating spacetime. This is due to the presence of momentum and acceleration perpendicular to the equatorial plane for this motion, similar to the problem of circular motion in Taub-NUT spacetime as described in \cite{Jefremov:2016dpi}. In light of this issue, we are interested in investigating whether an external magnetic field can enable stable circular motion in accelerating or Taub-NUT spacetime for a charged timelike object. We will address this problem in our future research.

\section*{Acknowledgement}

This work is supported by Kemendikbudristek.

\end{document}